\documentclass[aps,prd,twocolumn,superscriptaddress,nofootinbib,preprintnumbers]{revtex4-2}
\usepackage[dvipsnames]{xcolor}
\usepackage[colorlinks=true,linkcolor=MidnightBlue,citecolor=MidnightBlue,urlcolor=MidnightBlue]{hyperref}
\usepackage{graphicx}
\usepackage{amsmath,amssymb,amsfonts}
\usepackage{bm}
\usepackage{booktabs}
\usepackage{braket}
\usepackage{orcidlink}
\usepackage{multirow}
\usepackage{gensymb}
\usepackage{siunitx}
\usepackage{tikz}
\usetikzlibrary{arrows.meta}
\usepackage{aas_macros}

\newcommand{\Br}{\textrm{Br}}

\definecolor{myred}{HTML}{D55E00}
\definecolor{myorange}{HTML}{E69F00}
\definecolor{mygreen}{HTML}{009E73}
\definecolor{myblue}{HTML}{0072B2}

\begin{document}

\preprint{\tt   FERMILAB-PUB-26-0219-T}

\title{AXIS Could Have Accessed Dark Matter Decays}

\author{Joshua W. Foster\,\orcidlink{0000-0002-7399-2608}}
\affiliation{Department of Physics, University of Wisconsin--Madison, Madison, WI, USA}

\author{Inci Karaaslan\,\orcidlink{0000-0003-3784-9879}}
\affiliation{Department of Physics, University of Chicago, Chicago, IL, USA}

\author{Gordan Krnjaic\,\orcidlink{0000-0001-7420-9577}}
\affiliation{Theoretical Physics Division, Fermi National Accelerator Laboratory, Batavia, IL, USA}
\affiliation{Kavli Institute for Cosmological Physics, University of Chicago, Chicago, IL, USA}
\affiliation{Department of Astronomy \& Astrophysics, University of Chicago, Chicago, IL, USA}

\author{Nimrod Shapir\,\orcidlink{0009-0003-0278-2616}}
\affiliation{Department of Physics, University of Chicago, Chicago, IL, USA}
\date{\today}

\begin{abstract}
The Advanced X-ray Imaging Satellite (AXIS) was a mission concept designed to improve upon the sensitivity and spatial resolution of the Chandra X-ray Observatory and XMM-Newton. Although AXIS was not selected for implementation, the concept study defined a mature instrument design with low-background, arcsecond imaging over the $0.3$--$10$ keV energy range, a large effective area, and a wide field of view. These capabilities provide a useful benchmark for the dark matter decay sensitivity of future X-ray observatories. We estimate the reach of a future AXIS-like instrument for narrow photon lines from decaying keV-scale dark matter, including axion-like particles and sterile neutrinos. For Galactic center observations, we find projected lifetime sensitivities of order $10^{31}$ s, improving upon existing limits by up to an order of magnitude over part of the keV mass range.
\end{abstract}
\maketitle

\section{Introduction}
Dark matter (DM) is a fundamental constituent of the universe, with an enormous body of evidence supporting its existence across astrophysical and cosmological scales (see \cite{Bertone:2016nfn} for a review). However, while the available evidence is consistent with DM being a fundamental particle, it is unaccounted for in the Standard Model (SM) of particle physics, and its exact particle nature remains poorly understood. 

Indirect detection provides a powerful way to search for metastable DM whose decay products include SM particles. In the keV mass range, two-body decays can produce narrow X-ray lines that appear as excesses over astrophysical and instrumental backgrounds. This possibility is realized in well-motivated scenarios including axion-like particles (ALPs) \cite{Adams:2022pbo} and sterile neutrinos \cite{Abazajian:2017tcc}. Future X-ray observatories with large effective areas, wide fields of view, low backgrounds, and excellent angular resolution therefore provide a promising route to improved sensitivity to decaying DM.

The Advanced X-ray Imaging Satellite (AXIS) was developed as a next-generation X-ray mission concept intended to surpass the capabilities of the Chandra X-ray Observatory and XMM-Newton \cite{Mushotzky:2019lpm, Reynolds:2023vvf}. AXIS was designed with a 24' diameter angular field of view (FOV) and was optimized for soft X-ray observations over the $0.3$--$10$ keV energy range~\cite{Reynolds:2023vvf}. Although AXIS was not selected for implementation, its instrument concept can provide a concrete benchmark for the DM reach of future X-ray missions, a possibility that was discussed in~\cite{koss2026advancedxrayimagingsatellite}. In particular, the large effective area, low background, and arcsecond-scale imaging of a future observatory with similar instrument specifications (described as ``AXIS-like'' hereafter) are well-suited to searches for faint X-ray lines from DM decays.

\begin{figure}[!t]
    \centering
    \includegraphics[width=0.48\textwidth]{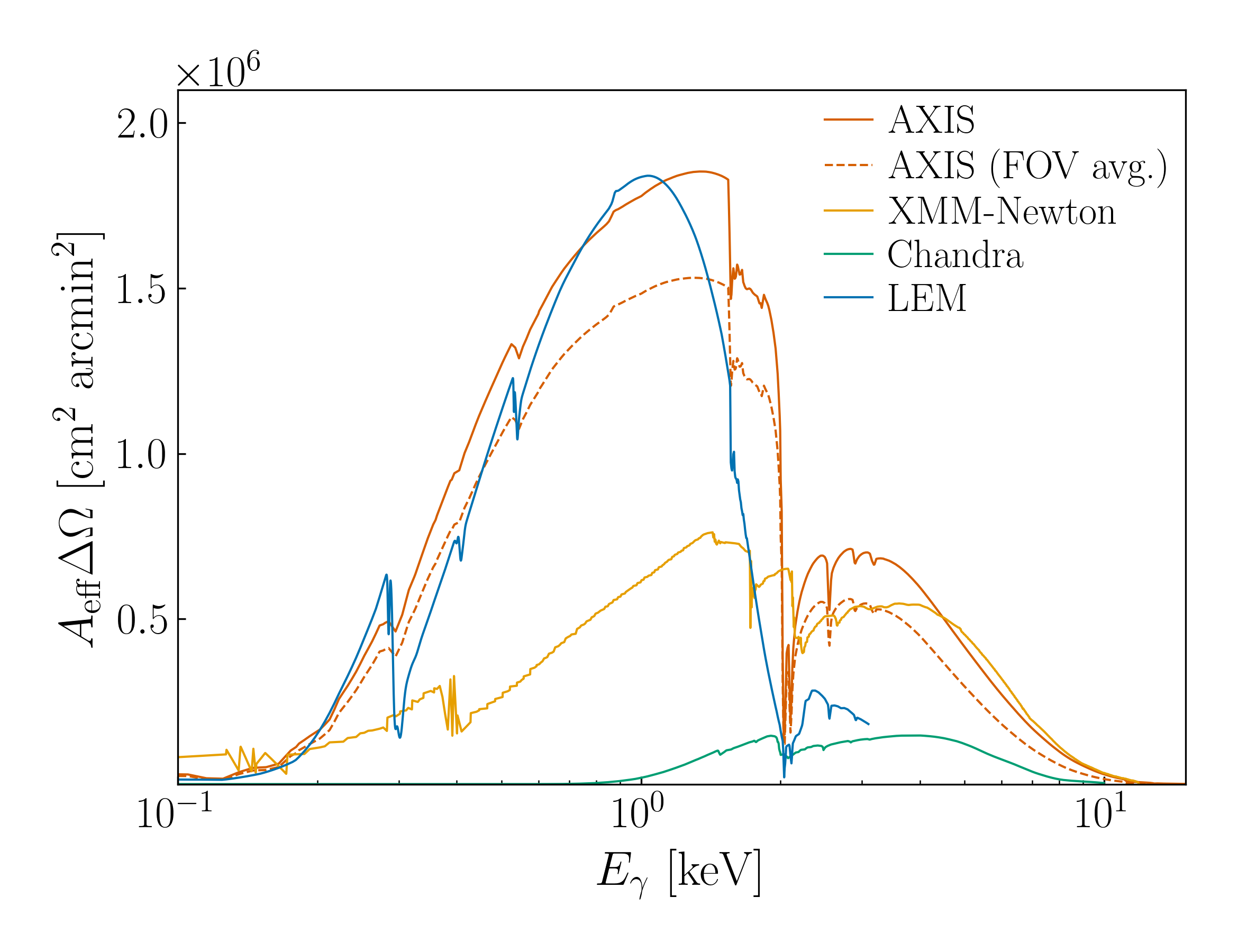}
    \caption{Projected on-axis grasp of AXIS \cite{axis_simulation_resources} (solid red curve), compared to XMM-Newton \cite{esa_xmm_uhb} (orange), Chandra \cite{chandra} (green) and LEM \cite{Kraft:2022mnh} (blue). The FOV-averaged grasp of AXIS \cite{axis_simulation_resources} is also included (dashed red curve).}
    \label{fig:grasp}
\end{figure}

The combination of large effective area and angular FOV gives AXIS an extensive grasp, $A_{\rm eff}\Delta\Omega$, of approximately $\SI{1.6e6}{cm^2.arcmin^2}$ at 1 keV (FOV averaged). The grasp is compared in Fig.~\ref{fig:grasp} with that of XMM-Newton, Chandra, and the Line Emission Mapper (LEM), another proposed next-generation X-ray telescope. AXIS's optimized mirror design also yields a point spread function (PSF) that is smaller by a factor of 6 compared to Chandra and by a factor of 10 compared to XMM-Newton~\cite{Mushotzky:2019lpm,axis_simulation_resources}. From the perspective of DM detection, the large photon collection rate and improved angular resolution would enable searches for fainter lines from DM decay while helping to separate signal photons from astrophysical backgrounds~\cite{AXIS}.

The parametric reach of such an instrument can be understood from a simple scaling estimate. For DM particles of mass $m$ decaying to nearly monochromatic photons of energy $E_\gamma \simeq m/2$, let $N_{\rm bg}$ denote the background counts in the energy range over which the line signal is distributed by the detector response. An AXIS-like telescope, then, would be sensitive at approximately $2\sigma$ statistical significance to lifetimes of order
\begin{equation}
\!\!
\tau \! \sim \SI{e31}{s} \bigg(\frac{\rm keV}{m}\bigg) 
\!\bigg(\!\frac{t_{\rm obs}}
{\SI{e8}{s}}\!\bigg)\! \bigg(\!\frac{A_{\rm eff}}{\SI{e3}{cm^2}}\!\bigg) 
\!\bigg(\!\frac{\Delta\Omega}{\SI{e-5}{sr}}\!\bigg) \! \sqrt{\frac{10^{6}}{N_{\rm bg}}},
\end{equation}
where $A_{\rm eff}$ is the effective area, $t_{\rm obs}$ is the exposure, and $\Delta \Omega$ is the angular FOV of the telescope. This scaling illustrates why the combination of large collecting area, wide FOV, long exposure, and low background is central to the projected sensitivity.

In this paper, we study the sensitivity of AXIS to DM decay channels that produce observable photon lines in order to inform the capabilities of a future AXIS-like telescope. We focus on observations of the Galactic center (GC), and include both the Galactic decay signal and the isotropic extragalactic contribution accumulated over cosmic redshifts. This paper is organized as follows. In Sec.~\ref{sec:2}, we describe the DM decay signal model, including the Galactic DM halo profiles and extragalactic contribution. Section~\ref{sec:3} details the signal and background modeling for AXIS projections. In Sec.~\ref{sec:4}, we present projected sensitivities first in terms of a model-agnostic DM lifetime and then reinterpreted in the context of ALPs and sterile neutrinos, respectively. We summarize our findings in Sec.~\ref{sec:5}.

\section{Modeling Dark Matter Decay Flux}\label{sec:2}
\begin{figure}[t]
\centering
\begin{tikzpicture}[scale=0.65]

\draw[dashed] (0,0) ellipse (6 and 2);

\draw[thick] (-6,0) -- (6,0);

\coordinate (Sun) at (-4,0);
\coordinate (GC) at (0,0);
\coordinate (P) at (2.5,1.4);
\coordinate (A) at (-1.5,0);
\coordinate (C) at (-1.5,0.2);
\coordinate (T) at (2.5,2.8);

\draw (Sun) -- (GC);
\draw (Sun) -- (P); 

\draw[gray, very thin] (2.3, 1.35) -- (2.3, 1.55);
\draw[gray, very thin] (2.3, 1.55) -- (2.5, 1.59);

\draw[mygreen,thick] (Sun) -- node[above] {$s$} (T);
\draw[myorange,thick] (GC) -- node[left] {$r$} (T);

\draw[myred,thick] (GC) -- node[below] {$R$} (P);
\draw[myblue,thick] (P) -- node[right] {$z$} (T);

\fill[black] (GC) circle (2pt) node[black,below] {GC};
\fill[black] (T) circle (2pt) node[black,above] {Target};
\fill[black] (Sun) circle (2pt) node[black,below] {Sun};


\draw[-{Stealth[length=1mm]}] (A) ++(-0.35,0.46) arc (12:23:2.2);
\node at (-1.6,0.75) {$b$};

\draw[-{Stealth[length=1mm]}] (A) ++(0.3,0) arc (0:12:2.85);
\node at (-0.95,0.35) {$\ell$};

\end{tikzpicture}

\caption{Line-of-sight geometry in Galactic coordinates. $s$ and $r$ denote the line-of-sight and galactocentric distances. $R$ and $z$ represent cylindrical coordinates with respect to the GC, while $\ell$ and $b$ are the Galactic longitude and latitude, respectively.}
\label{fig:gal_coords}
\end{figure}
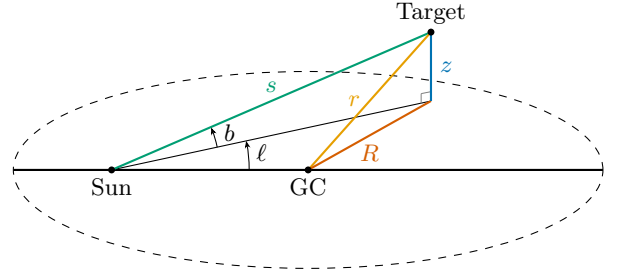

We begin by modeling the photon flux from decaying DM before convolving it with the instrumental response. For Galactic decays, the flux integrated over an angular region $\Delta \Omega$ can be written as~\cite{Hooper:2018kfv}
\begin{equation}
    \frac{d\phi}{dE_\gamma}
    =
    \frac{1}{4\pi m \tau}
    \frac{dN_\gamma}{dE_\gamma}
    \int_{\Delta \Omega} d \Omega
    \int_{\rm LOS} ds \, \rho(s, \hat\Omega)~,
\end{equation}
where $m$ is the DM mass, $\tau$ is its lifetime, $dN_\gamma/dE_\gamma$ is the photon energy distribution per decay, and $\rho(s,\hat\Omega)$ is the DM density evaluated at distance $s$ from the Sun along the line of sight $\hat\Omega$. It is useful to define the decay $D$-factor~\cite{Rodd:2018zrb, Cirelli:2010xx}
\begin{equation}
    D(\hat\Omega)
    \equiv
    \int_{\rm LOS} ds\,\rho(s,\hat\Omega),
\end{equation}
so that the Galactic decay flux can be written as
\begin{equation}
    \frac{d\phi}{dE_\gamma}
    =
    \frac{1}{4\pi m \tau}
    \frac{dN_\gamma}{dE_\gamma}
    \int_{\Delta \Omega} d \Omega \, D(\hat\Omega).
    \label{eq:FluxDMRestFrame}
\end{equation}
In this work, we take the distance of the Sun from the GC to be $r_\odot = 8.275$ kpc~\cite{ParticleDataGroup:2024cfk}. For the spherically symmetric halo profiles considered below, the density depends only on the galactocentric distance $r$, which is related to the line-of-sight coordinate by
\begin{align}
    r^2 = r_\odot^2 + s^2 - 2 r_\odot s \cos b \cos \ell,
\end{align}
where $\ell$ and $b$ are the Galactic longitude and latitude, respectively, as illustrated in Fig.~\ref{fig:gal_coords}. The differential solid angle is $d\Omega = \cos b\,d\ell\,db$, with $\ell$ and $b$ measured in radians.

For our model-agnostic analysis, we assume that a DM particle $\chi$ decays to two photons. The final-state photons are emitted monochromatically in the DM rest frame with energy $E_\gamma = m/2$, so the differential photon decay spectrum is
\begin{align}
    \frac{dN_\gamma}{dE_\gamma} = 2 \delta\left(E_\gamma - \frac{m}{2}\right).
    \label{eq:dNdERest}
\end{align}
When we later consider sterile neutrino DM, the relevant decay is instead $\nu_s\to\nu\gamma$ where only one photon is produced per decay, and hence we omit the overall factor of 2 in Eq.~\eqref{eq:dNdERest}. For keV-scale sterile neutrinos, the active neutrino mass is negligible, so the monochromatic decay spectrum approximation remains valid.

A decay line observed by an X-ray telescope will be broadened by the velocity dispersion of DM in the Milky Way halo, with characteristic velocities $v\sim10^{-3}$ producing an intrinsic relative width $\Delta E/E\sim10^{-3}$. By comparison, the energy resolution of AXIS is $\sim70$ eV at 1 keV, with relative resolution of order $\Delta E/E\sim 10\%$ across the range of energies relevant to this work~\cite{Mushotzky:2019lpm, Reynolds:2023vvf}. We therefore neglect Doppler broadening and treat the signal as monochromatic in the observer frame prior to convolution with the instrumental response. The observed signal is then obtained by integrating the intrinsic spectrum over the telescope bandpass and forward-modeling it through the response matrix, as described in Sec.~\ref{sec:3}. For instruments with substantially better energy resolution, however, accurate modeling of the astrophysically broadened decay line may be important; see Refs.~\cite{Kraft:2022mnh, Dessert:2023vyl, Krnjaic:2023odw}.

The total DM decay signal has two components: a Galactic contribution, sourced by the Milky Way halo, and an extragalactic contribution, accumulated over cosmological history. The Galactic contribution depends on the assumed DM density profile through the line-of-sight integral in Eq.~\eqref{eq:FluxDMRestFrame}. The extragalactic contribution is approximately isotropic before Galactic absorption and appears as a broadened continuum because photons emitted at higher redshift are observed at lower energy. We describe these two contributions in turn, followed by our treatment of absorption.

\subsection{Galactic Dark Matter Halo Profiles}\label{sec:profiles}

The distribution of DM in the Milky Way halo is an important input for the Galactic decay flux. Since the inner slope of the halo profile remains uncertain \cite{Benito:2019ngh}, we consider two spherically symmetric benchmark profiles whose densities decrease with galactocentric distance \cite{PerezdelosHeros:2020qyt}.

We first consider the Navarro-Frenk-White (NFW) profile, given by \cite{Navarro:1996gj}
\begin{align}
\rho_{\rm NFW}(r) = \dfrac{ \rho_0 }{ (r/r_s) \left( 1 + r/r_s \right)^2 },
\label{eq:nfw}
\end{align}
where $r_s=19.1\,\mathrm{kpc}$ is the scale radius~\cite{Foster:2021ngm}. The normalization $\rho_0$ is chosen to recover the local DM density $\rho_\odot=\SI{0.29}{GeV.cm^{-3}}$ used in Ref.~\cite{Foster:2021ngm}, facilitating comparison with the analysis performed therein.

We also consider the Einasto profile, given by \cite{1965TrAlm...5...87E}
\begin{align}
\rho_{\rm Einasto}(r) =
\rho_0
\exp \left[ -(r/r_s)^a \right],
\label{eq:einasto}
\end{align}
with scale radius $r_s=3.86\,\mathrm{kpc}$ and shape parameter $a=0.91$~\cite{2024MNRAS.528..693O}. For the Einasto profile, we instead fix $\rho_0$ such that $\rho_\odot=\SI{0.43}{GeV.cm^{-3}}$, as used in Ref.~\cite{Krnjaic:2023odw}. Since the decay flux is directly proportional to the halo normalization, our results can be rescaled linearly for a different choice of the local DM density.

\subsection{Extragalactic Dark Matter Decay Flux}\label{sec:EGflux}

In addition to the Galactic contribution from the Milky Way halo, decays of cosmological DM produce an approximately isotropic extragalactic signal. Unlike the Galactic signal, which appears as a narrow line before convolution with the instrumental response, the extragalactic contribution is broadened by cosmological redshift. Photons emitted at redshift $z$ with energy $m/2$ are observed at energy $E_\gamma=m/[2(1+z)]$, so the observed extragalactic spectrum extends over $0<E_\gamma\leq m/2$ even when the rest-frame decay spectrum is monochromatic.

For the decay spectrum in Eq.~\eqref{eq:dNdERest}, the FOV-integrated extragalactic photon flux can be written as~\cite{Essig:2013goa}
\begin{align}
\frac{d \phi}{dE_\gamma}
=
\frac{\Omega_{\mathrm{DM}} \rho_c}{2 \pi H_0 m^2 \tau}
\sqrt{\frac{2 E_\gamma}{m}}
\frac{\Delta \Omega}{\sqrt{\Omega_m+\Omega_{\Lambda}(2 E_\gamma / m)^3}},
\label{eq:EG}
\end{align}
for $E_\gamma \leq m/2$, and vanishes above the endpoint. Here $\Delta\Omega$ is the instrumental FOV. We take $H_0 = 67.4$ km s$^{-1}$ Mpc$^{-1}$, $\rho_c = \SI{4.79e-6}{GeV.cm^{-3}}$, $\Omega_{\rm DM} = 0.264$, $\Omega_{\rm m} = 0.313$, and $\Omega_\Lambda = 1 - \Omega_m = 0.687$~\cite{ParticleDataGroup:2024cfk}, where $\Omega_i\equiv \rho_i/\rho_c$ denotes the fractional density of component $i$. This expression treats the extragalactic DM distribution as homogeneous on the angular scales relevant for our analysis and neglects small anisotropies from large-scale structure.

\subsection{Signal Attenuation}

Both the Galactic and extragalactic signals are attenuated at low energies by photoelectric absorption, with the effect becoming especially important below $E_\gamma\lesssim\SI{1}{keV}$ \cite{1983ApJ...270..119M}. We model the attenuation along a line of sight $\hat\Omega=(\ell,b)$ by multiplying the differential flux per solid angle by an energy- and direction-dependent absorption factor,
\begin{align}
\frac{d^2\phi_{\rm att}}{dE_\gamma d\Omega}(\hat\Omega,E_\gamma)
=
\frac{d^2\phi}{dE_\gamma d\Omega}(\hat\Omega,E_\gamma)
e^{-N_\mathrm{H}(\hat\Omega)\sigma(E_\gamma)}.
\label{eq:attenuation}
\end{align}
The attenuated flux integrated over the instrumental FOV is then
\begin{align}
\frac{d\phi_{\rm att}}{dE_\gamma}
=
\int_{\Delta\Omega} d\Omega\,
\frac{d^2\phi_{\rm att}}{dE_\gamma d\Omega}(\hat\Omega,E_\gamma).
\end{align}
Here $N_\mathrm{H}$ is the neutral atomic hydrogen column density, obtained from Ref.~\cite{2016A&A...594A.116H}, and $\sigma(E_\gamma)$ is the photoelectric absorption cross section per hydrogen atom, obtained from Ref.~\cite{1983ApJ...270..119M}. We use $N_\mathrm{H}$ as a proxy for the total gas column density along the line of sight, since it is directly measured through 21 cm emission and captures the dominant spatial variation in the soft X-ray attenuation.

For the extragalactic component, this prescription accounts only for absorption after the photons enter the Milky Way. We do not model additional attenuation accumulated along the extragalactic line of sight, including absorption by the intergalactic medium or gas associated with intervening structures. Such effects could further suppress the extragalactic signal at low observed energies, so our treatment of extragalactic attenuation should be interpreted as approximate.

\section{Modeling X-ray Observations with an AXIS-like Telescope}\label{sec:3}

In this section, we describe how the DM decay fluxes from Sec.~\ref{sec:2} are converted into predicted photon counts for any AXIS-like X-ray observatory. This requires modeling the telescope response, the observing exposure, and the expected instrumental and astrophysical backgrounds within the FOV. These ingredients determine the energy-resolved signal and background counts used to derive projected sensitivities to DM decay.

The numerical values used to obtain the results in Sec.~\ref{sec:4} are based on publicly available AXIS simulation resources updated on 11 February 2025, including the AXIS instrument response files and effective area. The most recent version of the background models, in particular the non-X-ray background model, was provided to us via private communication~\cite{axis_private_backgrounds}.

\subsection{Instrument Response and Signal Counts}

The expected signal counts are obtained by forward modeling the incident DM decay flux through the AXIS instrument response. The effective area $A_\mathrm{eff}$, encoded in the ancillary response file (ARF), determines the photon collection rate as a function of input photon energy. The redistribution matrix file (RMF) then maps photons from input energy bins into detector energy bins, accounting for the finite energy resolution of the detector. Because the decay signal is diffuse, we use the FOV-averaged effective area when modeling the emission.

Taking the input-energy bin edges of the discretized instrumental response to be $E_j^\mathrm{in}$ and the detector-energy bin edges to be $E_i^\mathrm{out}$, the number of signal photons appearing in the $i^\textrm{th}$ detector energy bin is
\begin{align}
    N_i^{\rm sig}
    =
    t_\mathrm{obs}
    \sum_j
    {\rm RMF}_{ij}
    \int_{E_j^\mathrm{in}}^{E_{j+1}^\mathrm{in}} dE_\gamma\,
    A_\mathrm{eff}(E_\gamma)
    \frac{d\phi_{\rm att}}{dE_\gamma}.
    \label{eq:forwardmodel}
\end{align}
Here $t_\mathrm{obs}\sim\SI{1.1e8}{s}$ is our nominal mission exposure~\cite{Mushotzky:2019lpm}, and ${\rm RMF}_{ij}$ is the redistribution matrix element giving the probability for a photon in the $j^\textrm{th}$ input energy bin to be reconstructed in the $i^\textrm{th}$ detector energy bin. The quantity $d\phi_{\rm att}/dE_\gamma$ is the attenuated flux integrated over the instrumental FOV, as defined in Sec.~\ref{sec:2}. For each hypothesized DM mass, the decay spectrum determines the incident photon energies and therefore fixes the predicted detector-level signal counts $N_i^{\rm sig}(m,\tau)$ up to the overall normalization set by the lifetime $\tau$.

\subsection{Background Models and Counts}\label{sec:backgrounds}

Quantifying the sensitivity of an AXIS-like telescope to X-ray signals from decaying DM requires us to estimate the expected background counts within the AXIS instrumental FOV. We include three background components: the soft X-ray background (SXB), the cosmic X-ray background (CXB), and the non-X-ray background (NXB). The SXB is predominantly due to thermal emission from diffuse hot gas in the Milky Way interstellar medium, including the Local Bubble and Galactic halo \cite{McCammon:2002gb}. The CXB stems primarily from distant active galactic nuclei \cite{Mushotzky:2000zw}. The NXB consists of particle-induced events and instrumental noise unrelated to celestial X-ray emission, and is the dominant background component for the analysis presented here.

The NXB spectrum provided by the AXIS collaboration is based on the Suzaku telescope and has been updated to reflect a mission design change from low Earth orbit to L2. The SXB model is based on data from HaloSat \cite{2022ApJ...936...72B} and XQC \cite{McCammon:2002gb}, using the median surface brightness from XMM-Newton for the hot halo \cite{Henley:2013swt}. Both the SXB and NXB data files have already been forward modeled through the AXIS instrument response and are therefore provided in units of counts.

The AXIS simulation files do not include the CXB, so we model its unabsorbed intensity as a power law with photon spectral index $\Gamma=1.45$ and 1 keV normalization $C=\SI{10.91}{photons.cm^{-2}.s^{-1}.sr^{-1}.keV^{-1}}$ \cite{Cappelluti:2017miu}. Unlike the SXB and NXB files, the CXB is modeled as an incident sky flux. We therefore apply the attenuation prescription of Eq.~\eqref{eq:attenuation} and then forward model the attenuated CXB through the same instrument response used for the signal.

\begin{figure}
\centering
\includegraphics[width=0.48\textwidth]{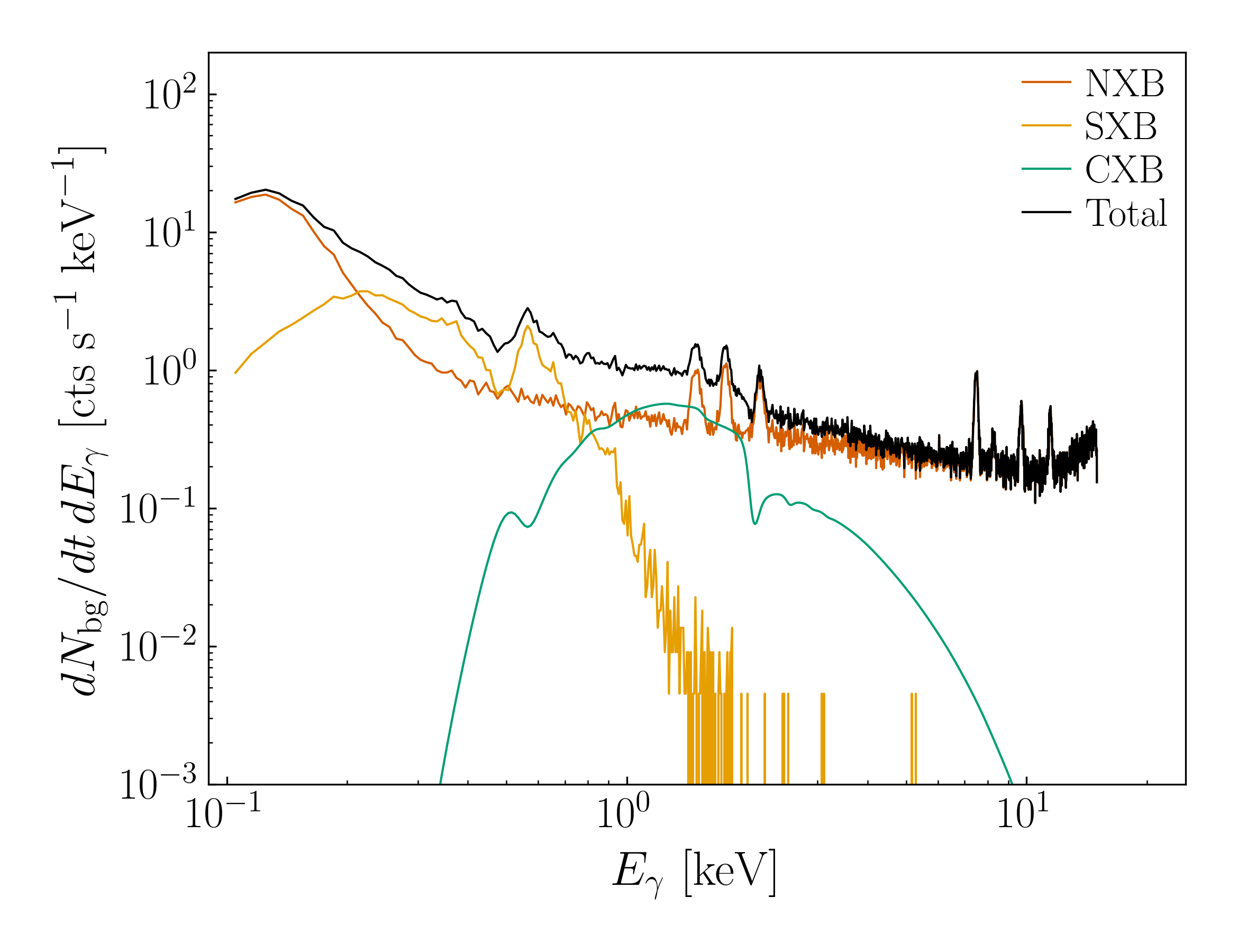}
\caption{Background model used in this analysis, including the NXB (red), SXB (orange), and CXB (green) components, as well as the total background (black). The CXB attenuation is evaluated at $(\ell,b)=(2\degree,2\degree)$ using Eq.~\eqref{eq:attenuation}. All components have been forward modeled through the AXIS instrument response.}
\label{fig:backgrounds}
\end{figure}

The total background counts in the $i^\textrm{th}$ detector energy bin are the sum of the components described above,
\begin{align}
N_i^{\rm bg}
=
N_i^{\rm SXB}
+
N_i^{\rm NXB}
+
N_i^{\rm CXB}.
\end{align}
Figure~\ref{fig:backgrounds} shows the background components and total background normalized per unit time and per unit energy.

\subsection{Sensitivity Estimates}

We determine projected sensitivities using an Asimov procedure in the weak-signal limit~\cite{Cowan:2010js}. In this approach, the data are taken to be equal to the expected background counts, and we ask what signal normalization would produce a specified expected detection significance. For the applications considered here, the expected signal is small compared to the background in each detector energy bin, $N_i^{\rm sig}\ll N_i^{\rm bg}$, so the Poisson likelihood-ratio test statistic reduces to
\begin{align}
    {\rm TS}(m,\tau)
    =
    \sum_{i=1}^{N_{\rm bin}}
    \frac{\big[N_i^{\rm sig}(m,\tau)\big]^2}{N_i^{\rm bg}}.
    \label{eq:teststatistic}
\end{align}
Here the sum runs over detector energy bins, and $N_i^{\rm sig}$ and $N_i^{\rm bg}$ are defined in the previous subsections. Equivalently, this corresponds to treating the uncertainty in each bin as dominated by Poisson fluctuations in the background counts, $\sigma_i=(N_i^{\rm bg})^{1/2}$, and summing the squared signal-to-noise ratio over bins.

In the asymptotic Gaussian limit, the expected detection significance for a positive line signal is $Z=\sqrt{{\rm TS}}$. We therefore define the projected $2\sigma$--$5\sigma$ sensitivity range by the lifetimes at which ${\rm TS}=4$ and ${\rm TS}=25$. Since $N_i^{\rm sig}\propto \tau^{-1}$, larger lifetimes produce smaller expected significance. We perform this procedure independently at each DM mass, using the corresponding decay spectrum and detector-level signal template. This treatment is purely statistical and does not include systematic uncertainties in the instrumental response, background modeling, or absorption prescription.

\section{Results}\label{sec:4}

\begin{figure}[t]
    \centering
    \includegraphics[width=0.48\textwidth]{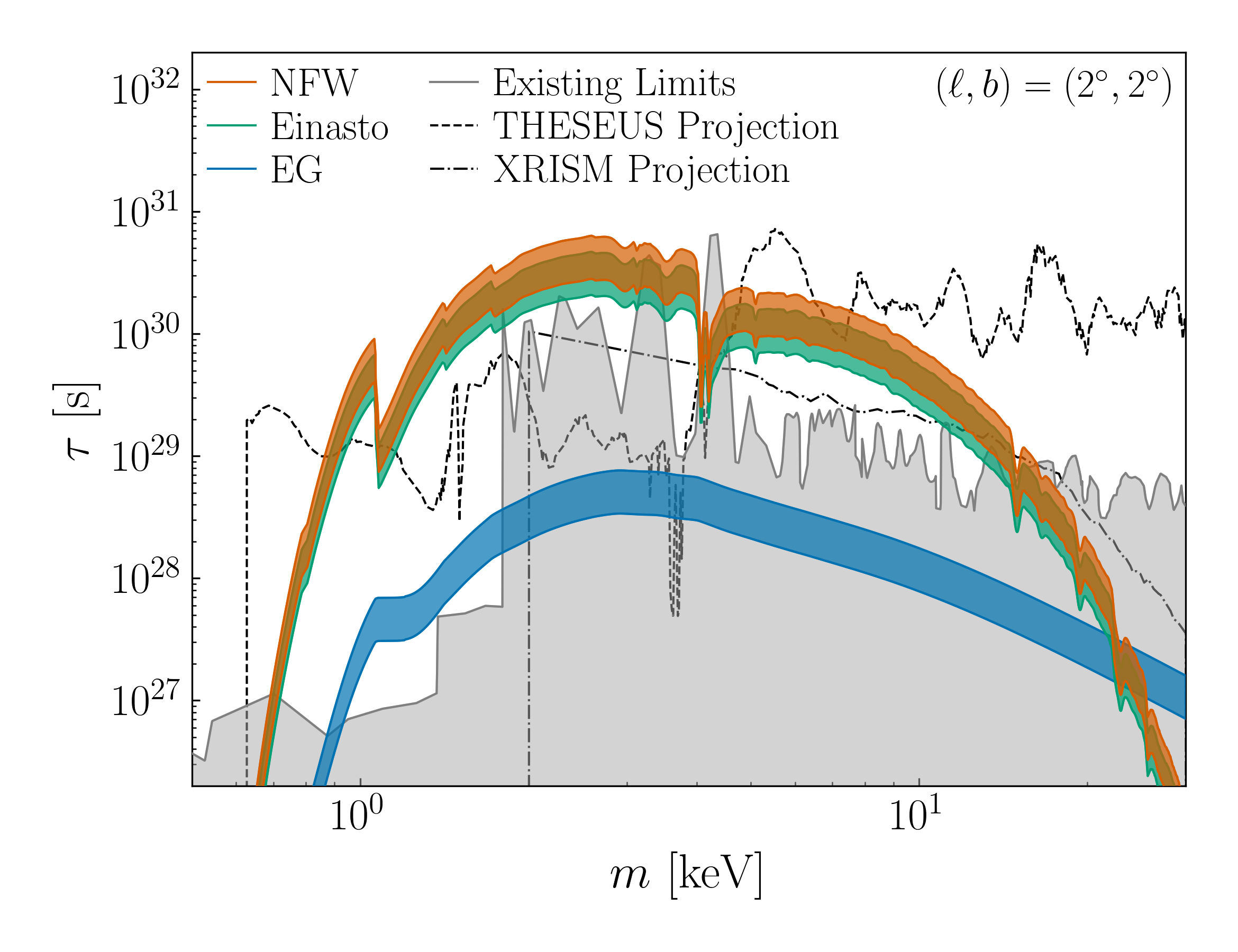}
    \caption{Projected sensitivity of AXIS to DM diphoton decays in the lifetime--mass plane, using the FOV-averaged grasp from Fig.~\ref{fig:grasp}, for $(\ell,b)=(2\degree,2\degree)$, to demonstrate the sensitivity reach of any AXIS-like telescope. The bands show the $2\sigma$--$5\sigma$ sensitivity range, defined by the lifetimes for which the expected detection statistic in Eq.~\eqref{eq:teststatistic} satisfies ${\rm TS}=4$ and ${\rm TS}=25$. The red and green bands assume only Galactic decays with the NFW and Einasto profiles, respectively, with parameters as described in Sec.~\ref{sec:profiles}. The blue band shows the corresponding projections for the extragalactic contribution, as discussed in Sec.~\ref{sec:EGflux}. Sharp features in the limit curves are due either to corresponding features in the effective area curve of the telescope (see Fig.~\ref{fig:grasp}), or to photoelectric absorption edges at particular energies. Existing limits on the DM mass and lifetime are taken from Ref.~\cite{AxionLimits}, and are primarily due to eROSITA~\cite{erosita}, XMM-Newton~\cite{Foster:2021ngm} and NuSTAR~\cite{nustar1, nustar2, nustar3}, while projected sensitivities for the future THESEUS and XRISM missions are shown separately. THESEUS projections are taken from Ref.~\cite{theseus_projection} and XRISM projections are taken from Ref.~\cite{xrism_projection}.}
    \label{fig:lifetimes}
\end{figure}

To estimate the lifetime reach of a future AXIS-like telescope, we choose a benchmark pointing within the GC, typically defined by $|\ell| \leq 5\degree$ and $|b| \leq 5\degree$~\cite{Rodd:2018zrb}. For the purpose of illustrating the full capability of such a telescope in the GC while avoiding uncertainties in the $N_\textrm{H}$ column density close to the exact center, as well as modeling uncertainties in the DM density distribution, we choose $(\ell,b) = (2\degree,2\degree)$. For a given survey strategy, the projected sensitivity is obtained by summing the test statistic over all pointings, with each field weighted by its exposure, background, and $D$-factor. However, in the absence of a specified survey strategy for AXIS, our fixed $(\ell,b)$ curves should be interpreted as benchmark single-field sensitivities rather than mission-averaged limits.

Figure~\ref{fig:lifetimes} shows the model-agnostic DM lifetime sensitivities (assuming diphoton decays) for the NFW and Einasto profiles. We observe that the projected sensitivities are up to an order of magnitude better than existing limits, which are primarily from eROSITA~\cite{erosita}, XMM-Newton~\cite{Foster:2021ngm}, and NuSTAR~\cite{nustar1, nustar2, nustar3} in the keV range.
\begin{figure*}[t]
    \centering
    \includegraphics[width=0.49\textwidth]{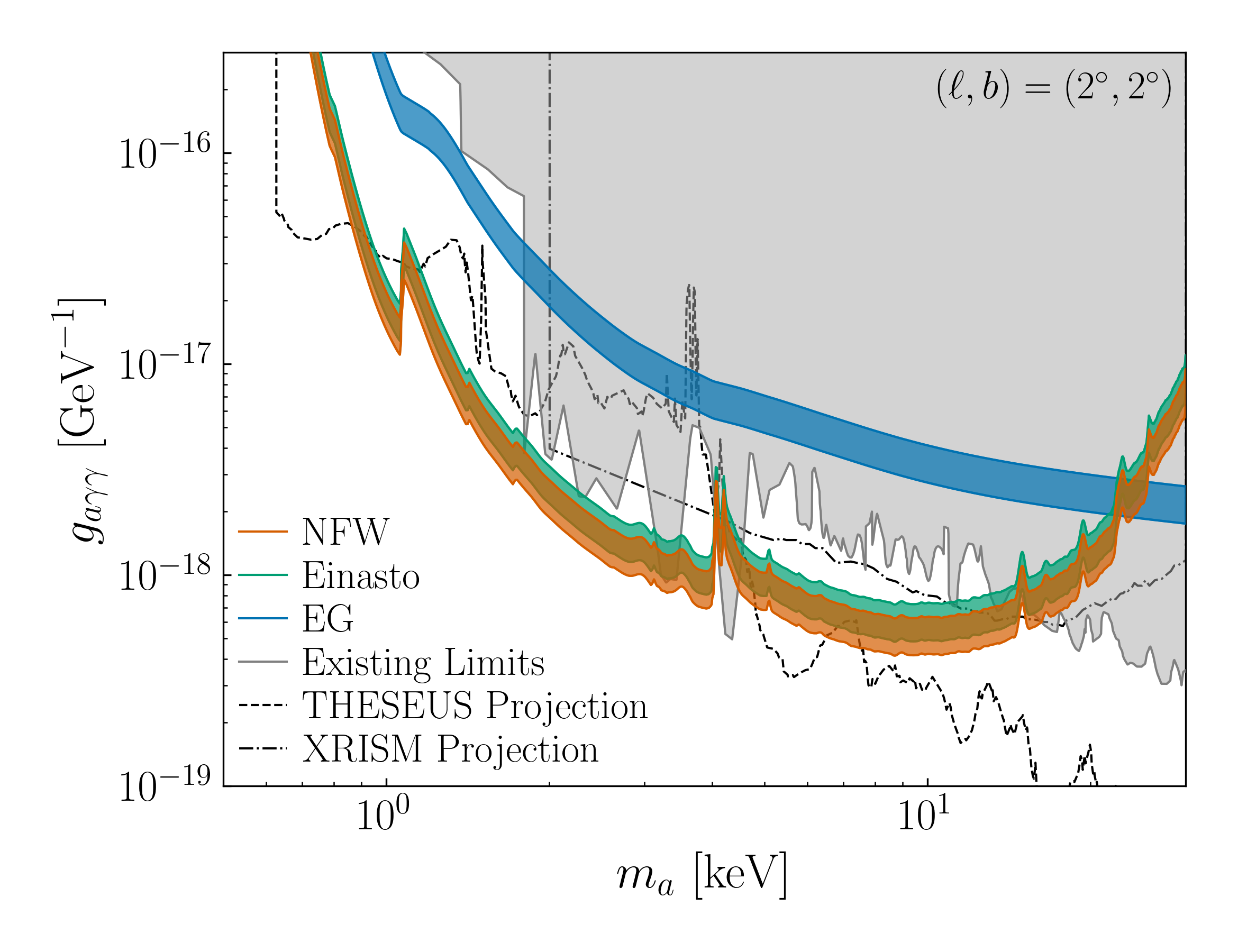}
    \includegraphics[width=0.49\textwidth]{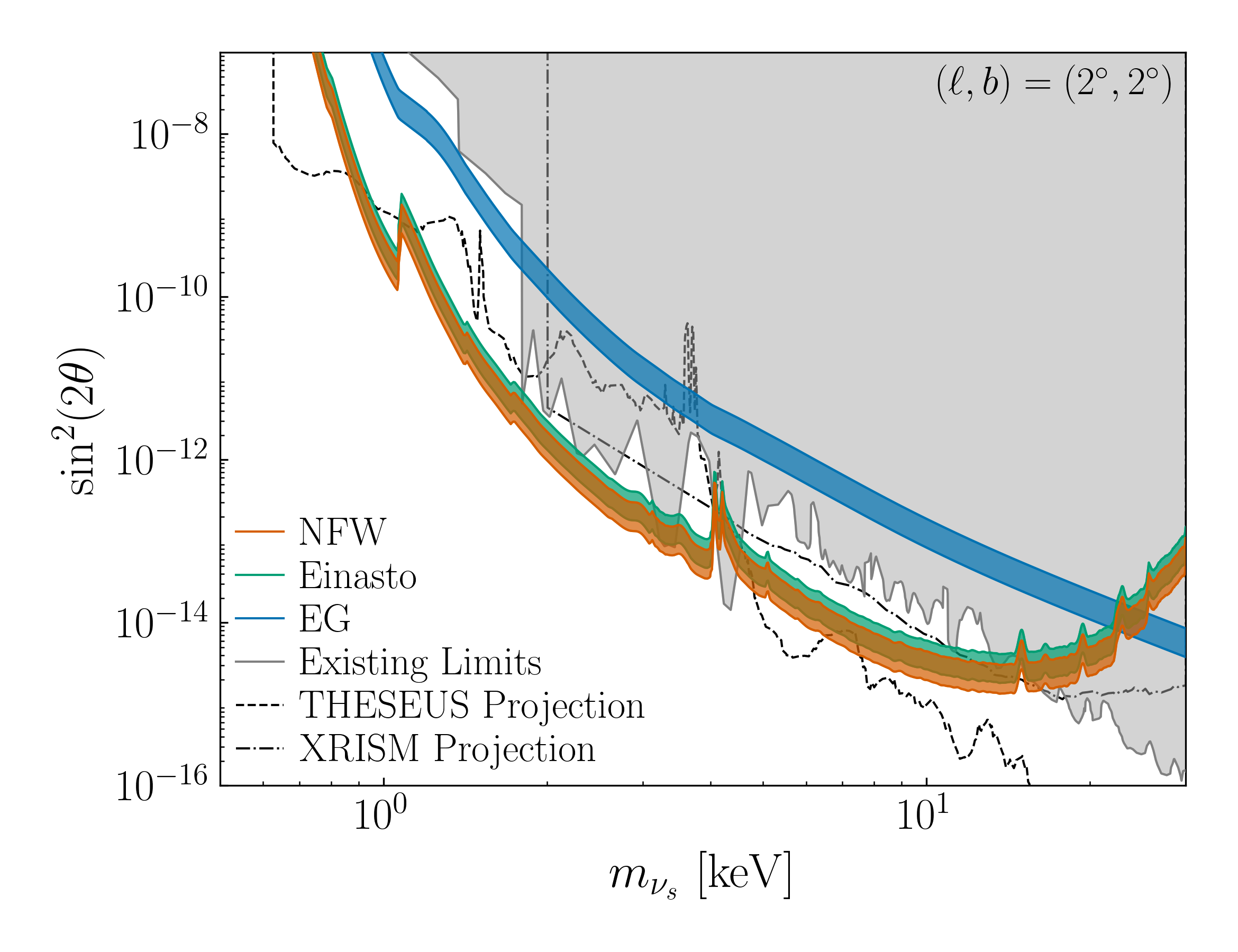}
    \caption{
    {\bf Left:}
    Projection for AXIS sensitivity to DM decays, now interpreted as a limit on the diphoton coupling $g_{a\gamma\gamma}$ of an ALP $a$ with mass $m_a$ and lifetime given by Eq.~\eqref{eq:ALP}.
    {\bf Right:} Projection for AXIS sensitivity to DM decays, now interpreted as a limit on the mixing angle $\theta$ of a sterile neutrino $\nu_s$ with mass $m_{\nu_s}$ and lifetime from Eq.~\eqref{eq:sterile}.}
    \label{fig:DMmodels} 
\end{figure*}

Comparing our results with a recent similar analysis for LEM \cite{Krnjaic:2023odw} (Fig.~1 therein), we find that the projected AXIS limits are about an order of magnitude weaker than the projected LEM limits. This difference between the two proposed observatories can be attributed to LEM's superior energy resolution ($\sim2$~eV~\cite{Kraft:2022mnh} as opposed to $\sim70$~eV for AXIS~\cite{Reynolds:2023vvf}). However, it should be noted that the analysis in~\cite{Krnjaic:2023odw} did not account for attenuation due to photoelectric absorption, which would diminish the sensitivity below $\sim1$ keV.

We now reinterpret our results in the context of specific types of DM models. We first consider the case of an ALP $a$ that decays to two photons, namely $a \rightarrow \gamma \gamma$. Assuming this is the dominant decay mode, the ALP lifetime can be written \cite{Cadamuro:2011fd}
\begin{equation}
    \tau = \frac{64\pi}{g^2_{a\gamma\gamma}m^3_a} \sim \SI{e30}{s} \left(\frac{\SI{e-17}{GeV^{-1}}}{g_{a\gamma\gamma}}\right)^2 \left(\frac{\rm keV}{m_a}\right)^3, \label{eq:ALP}
\end{equation}
where $g_{a\gamma\gamma}$ is the diphoton coupling. The left panel of Fig.~\ref{fig:DMmodels} shows the projected limits set by AXIS in the diphoton coupling vs. ALP mass plane.

We also consider the case of a sterile neutrino $\nu_s$ decaying into an active neutrino via $\nu_s \rightarrow \nu \gamma$, which produces only one photon per decay, so the factor of 2 in Eq.~\eqref{eq:dNdERest} is omitted. In this case, we can relate the DM lifetime to the active-sterile mixing angle $\theta$ through $\tau = \Br(\nu_s \to \nu \gamma)/\Gamma_\gamma$, where $\Gamma_\gamma$ is the sterile neutrino radiative width to photon lines \cite{Palazzo:2007gz}
\begin{equation}
\begin{split}
    \Gamma_\gamma&=\frac{9\alpha G_F^2\sin^2(2\theta)m_{\nu_s}^5}{1024\pi^4} \label{eq:sterile} \\
    &\sim\SI{e-32}{s^{-1}} \bigg(\frac{\sin^2(2\theta)}{10^{-10}}\bigg) \bigg(\frac{m_{\nu_s}}{\rm keV}\bigg)^5, 
\end{split}
\end{equation}
with $\alpha$ and $G_F$ the fine structure constant and the Fermi constant, respectively. We conservatively take $\Br(\nu_s\to\nu\gamma)\sim10^{-2}$ by comparing with the dominant decay channel $\nu_s\to 3\nu$. The right panel of Fig.~\ref{fig:DMmodels} shows the projected limits set by AXIS in the mixing vs. sterile neutrino mass plane. It should be noted that sterile neutrinos below $m_{\nu_s}\lesssim\SI{164}{eV}$ \cite{Boyarsky:2008ju} are subject to the Tremaine-Gunn bound \cite{Tremaine:1979we}, which excludes them from constituting the entirety of the DM content. Moreover, there are also model-dependent limits from structure formation that prohibit $m_{\nu_s} \lesssim$ a few keV if sterile neutrinos are produced through SM interactions in the early Universe \cite{Krnjaic:2023odw, DES:2020fxi}.

\section{Summary} \label{sec:5}

In this paper, we have estimated the sensitivity of an AXIS-like X-ray observatory to narrow photon lines from decaying dark matter. Using publicly available AXIS response files and background models, supplemented by a simple model for the cosmic X-ray background, we find that an AXIS-like instrument could improve upon existing limits on dark matter photon-line decays by up to an order of magnitude over part of the keV mass range. For Galactic center observations, the projected reach extends to lifetimes of order $\tau\sim\SI{e31}{s}$, depending on the assumed dark matter density profile and the treatment of absorption.

Although AXIS was not selected for implementation, its mature instrument concept provides a useful benchmark for assessing the dark matter discovery potential of future X-ray missions. The combination of large effective area, wide field of view, low background, and arcsecond-scale imaging would make an AXIS-like observatory well suited to searches for faint photon lines from dark matter decay. As discussed in Sec.~\ref{sec:4}, other proposed X-ray missions, such as the Line Emission Mapper, may achieve comparable or stronger lifetime sensitivities, particularly because of their improved energy resolution. Together, these results highlight the broader potential of future X-ray observatories to probe dark matter lifetimes in regimes that are beyond current observational reach.

\section*{Acknowledgments}
We thank Chris Dessert and Elena Pinetti for helpful conversations and particularly thank Tansu Daylan for providing us with the most recent version of the AXIS simulation files. This manuscript has been authored in part by Fermi Forward Discovery Group, LLC under Contract No. 89243024CSC000002 with the U.S. Department of Energy, Office of Science, Office of High Energy Physics.

\bibliographystyle{utphys3}
\bibliography{biblio}

\end{document}